\documentclass[11pt]{article}
\usepackage[margin=1in]{geometry}
\usepackage{lmodern}
\usepackage{sidecap}
\usepackage{xspace}
\usepackage{doi}
\usepackage{hyperref} 
\usepackage{tikz}
\usepackage{pgfgantt}
\usepackage{graphicx,pbox}
\usepackage{xcolor,hyperref,colortbl}
\usepackage{wrapfig}
\usepackage{enumitem}
\usepackage{gensymb}
\usepackage{comment}
\usepackage{lineno}
\newcommand\pubnumber{}
\newcommand\pubdate{\today}
\newcommand\pubblock{\rightline{\begin{tabular}{l} \pubnumber\\
          \pubdate \end{tabular}}}

\def\Title#1{\begin{center} {\Large #1 } \end{center}}
\def\Author#1{\begin{center}{ \sc #1} \end{center}}

\def\ee{$e^+e^-$}

\newcommand\snowmass{\begin{center}\rule[-0.2in]{\hsize}{0.01in}\\\rule{\hsize}{0.01in}\\
\vskip 0.1in Submitted to the  Proceedings of the US Community Study\\ 
on the Future of Particle Physics (Snowmass 2021)\\ 
\rule{\hsize}{0.01in}\\\rule[+0.2in]{\hsize}{0.01in} \end{center}}
\begin{document}
\begin{titlepage}
\snowmass
\pubblock

\Title{U.S. National Accelerator R\&D Program on Future Colliders
}

\Author{P.C.~Bhat$^{1,\dagger}$, S.~Belomestnykh$^{1,5}$, A.~Bross$^1$,  S.~Dasu$^6$, D.~Denisov$^4$, S.~Gourlay$^7$, S.~Jindariani$^1$, A.J.~Lankford$^{8,\dagger}$, S.~Nagaitsev$^{1,2,\dagger}$, E.A.~Nanni$^3$, M.A.~Palmer$^4$, T.~Raubenheimer$^3$, V.~Shiltsev$^1$, A.~Valishev$^1$, C.~Vernieri$^3$, F.~Zimmermann$^9$} 

\noindent
$^1${Fermi National Accelerator Laboratory} \\
$^2${University of Chicago }\\
$^3${SLAC National Accelerator Laboratory}\\
$^4${Brookhaven National Laboratory} \\
$^5${Stony Brook University}\\
$^6$ {University of Wisconsin, Madison} \\
$^7${Lawrence Berkeley National Laboratory, Retired}\\
$^8$ {University of California, Irvine} \\
$^9${CERN} \\

$^{\dagger}$ Lead Contacts; Email: pushpa@fnal.gov, andrew.lankford@uci.edu, nsergei@fnal.gov  \\ 


\end{titlepage}

\begin{abstract}
Future colliders are an essential component of a strategic vision for particle physics.  Conceptual studies and technical developments for several exciting future collider options are underway internationally.   In order to realize a future collider, a concerted accelerator R\&D program is required.   The U.S. HEP accelerator R\&D program currently has no direct effort in collider-specific R\&D area.  This shortcoming greatly compromises the U.S. leadership role in accelerator and particle physics.  In this white paper, we propose a new national accelerator R\&D program on future colliders and outline the important characteristics of such a program. 
\end{abstract}
\newpage
\tableofcontents
\newpage

\addcontentsline{toc}{section}{Executive Summary}

\section*{Executive Summary}

Future colliders are an essential component of a strategic vision for particle physics, for instance for the detailed study of the properties of the Higgs boson and for the exploration of higher mass scales than will be accessible at the HL-LHC.  In order to realize a future collider, a concerted accelerator R\&D program is required. 

While programs for the conceptual and technical developments for several future collider options are underway abroad, the U.S. accelerator R\&D program has a major gap in this area.  Currently, there is no part of the HEP program that supports early development of integrated collider concepts for strategic planning purposes, or that supports substantial engagement in the efforts on future colliders initiated abroad.  This gap compromises worldwide progress on promising concepts, such as the Future Circular Collider. It also inhibits early establishment of U.S. responsibilities in the design of facilities to be hosted outside the U.S., weakening the ability of the U.S. scientists and engineers to contribute in leadership roles. In addition, it impedes serious development of collider concepts that might be hosted in the U.S.  In short, this gap compromises the position of the U.S. as a leader in collider design and development and limits national aspirations for a U.S. leadership role in particle physics. 

To address this shortcoming in the U.S. HEP program, this white paper proposes a new national accelerator R\&D program on future colliders. \textbf{The overarching objective of the proposed U.S. national accelerator R\&D program on future colliders is to address in an integrated fashion the technical challenges of promising future collider concepts, particularly those aspects of accelerator design, technology, and beam physics that are not covered by the existing General Accelerator R\&D (GARD) program.}  This white paper outlines important characteristics of such a program.  The program will enable (a) synergistic U.S. engagement in ongoing global efforts (e.g., FCC, ILC, IMCC) and (b) developing collider concepts and proposals for options feasible to be hosted in the U.S.

\newpage

\section{Introduction} 

The first recommendation in the report ``Building for Discovery'' by the 2014 Particle Physics Project Prioritization Panel (P5) \cite{p52014}, which followed the previous U.S. Particle Physics Community ``Snowmass'' Study, was: ``Pursue the most important opportunities wherever they are, and host unique, world-class facilities that engage the global scientific community.'' The proposal for a national R\&D program on future colliders\cite{PCBhat_2022}, which we outline in this white paper, is in the spirit of that recommendation as the community builds its vision and strategy for the  particle physics program beyond the LHC and HL-LHC and the ongoing implementation of the LBNF/DUNE project.  The 2014 P5 report also described how colliders address three of P5's science drivers, ``Use the Higgs boson as a new tool for discovery'', ``Identify the new physics of dark matter'', and ``Explore the unknown: new particles, interactions, and physical principles''. In addition, it noted that ``the field relies on accelerators and instrumentation and on R\&D and test facilities for these technologies.''  The HEPAP Accelerator R\&D Subpanel, in its 2015 report on ``Accelerating Discovery'' also delivered several recommendations for rigorous R\&D in accelerator science and technology. 

As we celebrate the tenth anniversary of the discovery of the Higgs boson, the prevailing view of the worldwide particle physics community is that: (1) the next collider facility should be an $e^+e^-$  collider as a Higgs (and electroweak) factory \cite{alcc_HiggsFactory} to conduct detailed studies of the Higgs boson and exquisitely precise measurements of its couplings, and (2) a collider with energy reach of $\sim10$~TeV scale for new physics searches will be necessary in the longer term.
This view was expressed in the 2014 P5 report, it was stated explicitly in the 2020 Update of the European Strategy for Particle Physics, and it is reflected in summary report of Energy Frontier group of the 2021/22 Snowmass Community Planning Exercise.

Several options for each of these types of colliders have been under consideration globally, with various levels of technical readiness.  The leading candidates for a Higgs/EW Factory are (1) the International Linear Collider (ILC) in Japan, (2) the $e^+e^-$ Future Circular Collider (FCC-ee) at CERN, (3) Circular Electron-Positron Collider (CEPC) in China, and (4) Compact LInear Collider (CLIC) at CERN. Additional novel options for compact $e^+e^-$  colliders, such as Cool Copper Collider (C$^3$), high gradient ($\sim70$ MV/m) superconducting RF linear collider HELEN (High Energy LEptoN collider), and a Fermilab Site Filler circular $e^+e^-$ collider,  have emerged and are under investigation. For the higher energy frontier collider option, to probe the $\sim10$~TeV scale, the FCC-hh at CERN and SppC in China are the prime candidates proposed. Additional novel options are a collider-based on advanced acceleration concepts, for which active technology R\&D is being performed,  and a Muon Collider, which has recently garnered a lot of interest and excitement as an option. 

Given that the targeted R\&D and a long preparatory phase  will be needed before initiating the construction of such a collider, concerted efforts should be undertaken now to avoid major gaps in physics pursuits and to keep the expertise in the field intact.  It is also imperative that the United States remain a major player in the next global collider facility to be realized, whether it is built abroad or at home.  Therefore, it is crucial that a U.S. national R\&D program with a sharp focus on future colliders be launched by the DOE Office of High Energy Physics, in the earliest possible time frame.    We outline the goal, objectives and characteristics of such a national R\&D program in this white paper.   


\textbf{The overarching objective of the proposed U.S.~national accelerator R\&D program on future colliders is to address in an integrated fashion the technical challenges of promising future collider concepts, particularly those aspects of accelerator design, technology, and beam physics that are not covered by the existing General Accelerator R\&D (GARD) program.} 
This new program should be inclusive of future collider concepts, both for next generation colliders and for colliders further in the future, with priorities guided by the Particle Physics Project Prioritization Panel. It should be synergistic with international design studies and R\&D activities, as well as explore options for a  U.S. hosted collider.

The goal of the program is to inform decisions in down-selecting among the collider concepts by the next European strategy update and the next US community planning cycle, to help move towards realization of the next collider as soon as possible and subsequently to advance towards a collider at a higher energy scale. 
\section{Program Characteristics} 

The proposed program would have the following characteristics. These characteristics are conceived in order to best address the program’s goal and objectives. The characteristics are grouped by the proposed scope, organization, and coordination.

\subsection{Scope}

The program should address accelerator challenges, specific to the next colliders, for instance, for Higgs factories, and of concepts for colliders further in the future, for instance 10-TeV scale machines. It should complement the existing HEP GARD program, which presently contains generic long-term R\&D thrusts.

The program should be multifaceted in that it should not focus entirely on a single collider concept; nor should it be limited to certain R\&D thrusts. Nonetheless, the program should be selective among R\&D topics in order that its scope is not impractically expansive. Supporting multiple approaches in a synergistic way is not only cost-effective, it also increases the chances of success in converging on the most viable option and enhances the possibility of yielding additional technology benefits.

The selection of R\&D topics to be pursued by the program should be guided by the reports of the Particle Physics Project Prioritization Panel (P5). A mechanism will need to be established to perform the selection based on the reports of P5. For instance, guided by P5 and with the help of an advisory committee, the body managing the execution of the program could propose a selected set of R\&D activities that constitute a coherent program.

The program should integrate all critical accelerator R\&D needed to develop and optimize collider concepts. In this respect, the program should be much more than a collection of independent R\&D activities. It should include development of full collider concepts as well as R\&D on technologies that are not covered within the existing GARD thrusts. The existing DOE GARD program supports medium- to long-term beam physics and accelerator technology R\&D of interest for future machines across the broad field of potential accelerator applications that is supported by the DOE Office of Science. The proposed National Accelerator R\&D Program on Future Colliders would not overlap with GARD, but would make use of results achieved under GARD and apply them to various collider concepts. 
It should also address aspects and challenges of the accelerator design, technology, and beam physics specific to colliders that are not otherwise covered by GARD. Although the development of experiment concepts and detector R\&D are beyond the scope of the program proposed here, studies of the machine-detector interface are not.

Examples of R\&D topics that could be addressed are described in the white paper discussing future collider options \cite{PCBhat_2022} and several papers on individual collider concepts [~\cite{agapov2022future,aryshev2022international, nanni2022c,CCCstrategy,CCCfacility, belomestnykh2022higgs, MuCForumReport,stratakis2022muon}.
The budget for the national program will need to be worked out in the process of prioritizing and selecting R\&D projects. An impactful program might require an average annual investment of \$25M between now and the next Snowmass/P5 cycle.

\subsection{Organization}


The program should be a national collaborative effort open to U.S. national labs and universities interested in contributing to the R\&D activities defined by the program. Practical limitations on funds will require a process, e.g., a proposal/review process, by which funding is allocated. It is envisaged that many individual R\&D activities within the program will be conducted by collaborations among multiple institutions.

\subsection{Coordination}


In order to ensure that the program’s activities are integrated into a coherent program advancing developments and preparedness for future colliders, the program’s portfolio of R\&D activities should be centrally coordinated, and funding should be centrally allocated. A program management office could be hosted at a national lab.

The established coherent program should identify cross-cutting developments across subsets of collider concepts and foster R\&D activities that can be used in multiple collider concepts in order to optimally exploit synergies among various R\&D topics and thrusts and to make efficient use of resources. For new collider concepts that are in the early design phase, the program could provide integration and optimization prior to the formal conceptual design phase. For all collider concepts, the program could provide periodic assessment of coherence of activities and specifications. 

This national program should establish an appropriate level of coordination with collider design studies and R\&D outside the U.S., in order to avoid needless duplication of effort and in order to foster complementary studies.

To ensure continued alignment with P5 priorities and other important developments, the program should establish a mechanism for periodic assessment of program progress and direction. This mechanism could be provided by the aforesaid advisory committee composed of physics and accelerator experts.

\section{Summary}

The prevailing view of the global HEP community is that the next large collider facility should be an \ee collider as a Higgs/EW factory.   The physics case for such a collider is compelling, because it would enable the use of ``the Higgs boson as a new tool for discovery,'' as envisioned in the 2014 P5 report.  The community also recognizes that a collider beyond the capabilities of the HL-LHC, with energy reach to explore the $\sim10$~TeV scale, such as a $\sim100$~TeV hadron collider or a $\geq10$ TeV muon collider, will be necessary. 

In order to position the U.S. as a key player in these future HEP facilities,  whether hosted abroad or in the U.S., we have proposed in this white paper that the DOE Office of High Energy Physics launch an integrated national accelerator R\&D program focused on future colliders. We have outlined the potential scope of the program, and how it could be organized and coordinated.  The characteristics of each of these aspects have been discussed and can be summarized as follows:\\
\\
Scope:
\begin{itemize}
\setlength\itemsep{0.04em}
    \item Sharply focused on future colliders
    \item Spans accelerator design, technology and full concept development
    \item Complements the existing HEP GARD program
    \item Multifaceted but selective, and synergistic
    \item Integrates all critical R\&D for a concept
    \item Priorities guided by P5
\end{itemize}

\noindent
Organization:
\begin{itemize}
\setlength\itemsep{0.04em}
    \item Coherent national program
    \item Collaborative effort of U.S. national labs and universities
\end{itemize}

\noindent
Coordination:
\begin{itemize}
\setlength\itemsep{0.04em}
    \item Centrally coordinated and funded
    \item  Coordinated with global design studies and R\&D
    \item Periodic assessment 
\end{itemize}

This R\&D program would facilitate the realization of future collider facilities, thereby ensuring the continuation of the fruitful endeavors of HEP in advancing the frontiers of our knowledge of the universe. It will also ensure the critical recruitment, development, and retention of a skilled workforce in accelerator science and technology.

\section{Acknowledgments}
We gratefully acknowledge the support of the US Department of Energy and the US National Science Foundation.


\addcontentsline{toc}{section}{Bibliography}

\bibliographystyle{atlasnote}
\bibliography{bibliography.bib}

\end{document}